\documentclass[a4paper, fleqn, usenatbib]{mnras}
\usepackage{newtxtext,newtxmath}
\usepackage[T1]{fontenc}
\usepackage{ae,aecompl}
\usepackage{listings}
\usepackage{booktabs} 
\usepackage{multicol}
\usepackage{multirow}
\usepackage{afterpage}
\usepackage{graphicx}	
\usepackage{amsmath}	
\usepackage{amssymb}	
\usepackage[usenames,dvipsnames]{color}

\newcommand{\Msolar}{\mbox{\,$\rm M_{\odot}$}}        
\newcommand{\kmsec}{\mbox{\,$\rm km\,s^{-1}$}}        

	\newcommand{\pgstar}{\mbox{\,PG\,0909+276}}
	\newcommand{\uvstar}{\mbox{\,UVO\,0512--08}}
	\newcommand{\ledz}{\mbox{\,UVO\,0825+15}}
	\newcommand{\crimson}{\mbox{\,LS\,IV--14$^{\circ}$116}}

	\newcommand{\teff}{\mbox{\,$T_{\rm eff}$}}      
	\newcommand{\lgcs}{\mbox{\,$\log g / {\rm cm\,s^{-2}}$}}        
    \newcommand{\ebv}{\mbox{\,E(B-V)}}

	\newcommand{\nH}{\mbox{\,$n_{\rm H}$}}                 
	\newcommand{\nHe}{\mbox{\,$n_{\rm He}$}}               

\title[Chemical analysis of two hot subdwarfs]{Surface abundance and the hunt for stratification in chemically peculiar hot subdwarfs: PG\,0909+276 and UVO\,0512--08}
\author[J. F. Wild and C. S. Jeffery]{
	J. F. Wild$^{1,2}$ and
	C. S. Jeffery$^{1,3}$\\
	$^1$Armagh Observatory and Planetarium, College Hill, Armagh BT61 9DG, N. Ireland\\
	$^2$Department of Physics \& Astronomy, University of Sheffield, Sheffield, S3 7RH, UK\\
    $^3$School of Physics, Trinity College Dublin, Dublin 1, Republic of Ireland\\
}

\date{Received September 5, 2017}

\pubyear{2017}

\begin{document}
\label{firstpage}
\pagerange{\pageref{firstpage}--\pageref{lastpage}}
\maketitle

\begin{abstract}
	\citet{edelmann03th} identified two chemically peculiar hot subdwarfs, \pgstar\ and \uvstar, as having very high overabundances of iron-group elements. We obtained high-resolution ultraviolet spectroscopy in order to measure abundances of species not observable in the optical, and to seek evidence for chemical stratification in the photosphere.  Abundances were measured in three  wavelength regions; the optical 3900\AA-6900\AA\ range was re-analysed  to confirm consistency with \citet{edelmann03th}. Ultraviolet spectra were obtained  with the Space Telescope Imaging Spectrograph (STIS) on the Hubble Space Telescope (HST), covering the far-UV (1140\AA-1740\AA) and the near-UV (1740\AA-2500\AA). We computed a grid of theoretical LTE spectra to find basic parameters (effective temperatures, surface gravity, surface hydrogen and helium fractions). We measured abundances using a spectral-synthesis approach in each wavelength range. We confirm that several iron-group metals are highly enriched, including cobalt, copper and zinc, relative to typical sdB stars. We detect gallium, germanium, tin, and lead, similar to analysis of ultraviolet spectra of some other sdB stars. Our results confirm that \pgstar\ and \uvstar\ exhibit  peculiarities which make them distinct from both the normal H-rich sdB and intermediate He-rich sdB stars. The process which leads to this particular composition has still to be identified. 
\end{abstract}

\begin{keywords}
stars: abundances -- stars: fundamental parameters -- stars: chemically peculiar -- stars: individual (\pgstar) -- stars: individual (\uvstar) -- stars: subdwarfs
\end{keywords}


\section{Introduction}
	Hot subdwarfs form a class of compact, low-mass, evolved stars with very high surface temperatures. A subclass with spectral type B, having strong Balmer lines, weak neutral helium lines and no ionised helium lines, are known as sub-luminous B or sdB stars. These objects have masses $M < 0.5\Msolar$, and typically consist of a helium burning core surrounded by a thin, inert hydrogen layer that is unable to sustain fusion. They have effective temperatures (\teff) in the range 20\,000 to 35\,000\,K and surface gravities $\lgcs$ in the range 5.5 to 6.5, placing them on or near the blue end of the extreme horizontal branch on the Hertzsprung-Russell diagram \citep{heber16}. The majority are extremely helium-poor.  With \teff\ up to 45\,000\,K, hot subdwarfs showing both neutral and ionised helium lines are known as sdOB stars; they show a range of helium abundances from helium-weak though to extremely helium-strong. Most are located on or close to the helium main sequence.  
    
    The formation and evolution of hot subdwarfs is not fully understood, with one problem being the very large amount of mass loss required before the helium core ignites, and another being the number of subclasses observed. Several potential evolutionary tracks have been proposed, including a late helium flash in a post-giant star \citep{brown01,bertolami08}, common-envelope binary evolution \citep{han98,ivanova2013}, or a white dwarf merger \citep{iben90,saio02,zhang12a}. However, it is difficult to say to which track a given hot subdwarf belongs. 
	
	In a previous analysis of high-resolution optical spectra, two sdOB stars \pgstar\ and \uvstar\ were found to be hydrogen-weak, with peculiar super-solar abundances of metals with atomic number $Z\ge 7$ \citep{edelmann03th,geier13}. Bright enough to observe at low resolution with the International Ultraviolet Observer {\it IUE}, new high-resolution ultraviolet spectra have been obtained with the Hubble Space Telescope ({\it HST}) in order to better determine  the elemental composition of these two stars, including the identification of additional species. 

	It has been suggested in the analyses of other chemically peculiar, hydrogen-poor sdB stars \citep{naslim13} that extreme overabundances of heavy elements are primarily due to enrichment of the stellar atmosphere by selective radiative levitation of specific ions such that these ions form layers (strata) of high concentration in the line forming region of the photosphere. 
    
    Since different lines probe different physical depths depending on line strength and the local continuum, a secondary goal is to find evidence for chemical stratification within the photosphere. However, because the difference in line formation depths is only a tiny fraction of a typical sdB radius, this is a challenging goal.

\begin{table*}
\caption{Observations}
\label{t:obs}
\begin{tabular}{llllcrrrl}
\hline
Star & Instrument 				& Date 			& $R$ & $\lambda\lambda/$\AA &$n$ & $t_{\rm exp}/$s & S/N & Image \\
\hline
\pgstar\ & DSAZ FOCES			& 2000.01		& 30\,000 	& 3900 -- 6900 & 3 & 5400	& 30 	&  \\
&    {\it HST} STIS E140M 		& 2015.04.24	& 45\,800 	& 1140 -- 1740 & 1 & 2989 	& 20 	& OCKS02020 \\
&    {\it HST} STIS E230M 		& 2015.04.24	& 30\,000 	& 1740 -- 2500 & 1 & 2019 	& 50 	& OCKS02010 \\
&    {\it IUE} SWP LORES LAP 	& 1986.01.07 	& 260		& 1150 -- 2000 & 1 & 300 	&   	& SWP27469LL \\
&    {\it IUE} LWP LORES LAP 	& 1986.01.07 	& 320 		& 1850 -- 3350 & 1 & 450 	&  		& LWP07464LL \\
\hline
\uvstar\ & DSAZ FOCES			& 2000.01--02	& 30\,000 	& 3900 -- 6900 & 2 & 7200	& 30	&  \\
&    {\it HST} STIS E140M 		& 2015.02.25	& 45\,800 	& 1140 -- 1740 & 1 & 1006 	& 20 	& OCKS01010 \\
&    {\it HST} STIS E230M 		& 2015.02.25	& 30\,000 	& 1740 -- 2500 & 1 & 696 	& 30 	& OCKS01020 \\
&    {\it IUE} SWP LORES LAP 	& 1980.02.28 	& 260 		& 1150 -- 1950 & 1 & 235 	&  		& SWP08075LL \\
&    {\it IUE} LWR LORES LAP 	& 1980.02.28 	& 320 		& 1900 -- 3200 & 1 & 330 	&  		& LWR07043LL \\
\hline
\end{tabular}
\end{table*}

	Third, the question arises as to why these stars contain a substantially larger fraction of heavy metals than other similar hot subdwarfs \citep{otoole04}. A possible solution is that  heavier metals, due to their more complex electron configuration, have more spectral lines that could intercept and absorb a larger fraction of high-energy photons. This makes them more susceptible to radiative levitation than light ions with simpler outer electron structures. Evidence for chemical stratification in the photosphere could support this idea. 

    This paper summarizes the observations used (\S\ref{section:observations}), as well as the models and methods used to analyse the data (\S\ref{section:methods}). We determine the basic parameters and photospheric abundances for each star, in particular comparing results obtained in different wavelength regions (\S\ref{section:pg0909},\ref{section:uvo0512}). The results and their potential implications are discussed in the context of previously observed hot subdwarfs (\S\ref{section:discussion}). 

\section{Observations}
\label{section:observations}
	We use data from three sets of observations: high-resolution ultraviolet spectra from the Space Telescope Imaging Spectrograph (STIS), low-resolution ultraviolet spectrophotometry from {\it IUE}, and high-resolution optical spectra from the Fiber-Optics Cassegrain Echelle Spectrograph (FOCES) mounted on the 2.2m telescope of the Deutsch-Spanisches Astronomisches Zentrum at the Calar Alto Observatory, Spain. The latter were obtained and analysed by \citet{edelmann03th,geier13}. Details are given in Table~\ref{t:obs}. 
    
     For both stars, optical spectra show large sections that have either weak or no lines, although \citet{edelmann03th} notes that the optical spectrum of \uvstar\ contains many more lines than expected from a hot subdwarf. With several strong hydrogen and helium lines, and isolated lines of other species, the optical spectra provide the most robust means to determine basic parameters and to estimate abundances of some key elements.  
    
    Data were obtained with the {\it HST} in both the near ultraviolet (NUV) and far ultraviolet (FUV). Reduced data from the pipeline provided by Space Telescope Science Institute were obtained from the online archive. \pgstar\ was also observed with {\it HST}/STIS by Rauch in March 2017. These data are still proprietary at the time of writing.

    The FUV  is particularly crowded with blended absorption lines, with  neither star showing any regions that are clearly continuum. 
    Significant line blending makes normalization of this region a challenge, as it is unclear where the continuum should be set. In order to obtain a spectrum for which analysis was possible, it was assumed that the highest level fluxes observed  approximately represent the continuum and the spectrum was rectified accordingly. The normalization represents a potential source of systematic error which can be assessed when comparing results for different wavelength regions. 

	The NUV presents slightly fewer lines, as well as having some sections of continuum spectrum. This allows for much easier normalization. Additionally, the atomic data for  this region are  more complete than for the FUV. 
    
	We verified the radial and projected rotational velocities $v_{\rm rad}$, $v_{\rm rot} \sin i$ of each star from the optical spectra. 
     For \pgstar, we found $v_{\rm rad} = 20.7\kmsec$ and  $v_{\rm rot}\sin i < 2\kmsec$.
     For \uvstar, we found $v_{\rm rad} = 11.7\kmsec$ and  $v_{\rm rot}\sin i < 2\kmsec$.
    The sharp metal lines are consistent with a microturbulent velocity that is less than 2\kmsec\ in both stars.
     These values agree with results obtained using the same data by \cite{edelmann03th} and also reported by \citet{martin16}.

     We  compare our final models with spectra obtained with {\it IUE}, to confirm our measurements of the stars' \teff\ match historic data. 
    We use two of the four observations of \uvstar\ made by IUE on the $28^{th}$ of February 1980. Specifically, we use the data for which the star was observed at low dispersion, using the large aperture of the instrument. The first observation covers the far ultraviolet, and the second covers the near ultraviolet. \pgstar\ was observed with the same method, on the $7^{th}$ of January 1986.
    
    We also use photometric observations of both stars. We obtained 6 measurements of \pgstar\ in the B, V, R, J, H, and K bands \citep{hog00,zacharias09,cutri03} and 5 measurements of \uvstar\ in the B, V, J, H, and K bands \citep{hog00,cutri03}. These are considered in our fitting of the IUE data.

\section{Methods}
\label{section:methods}
	Model atmospheres and theoretical spectra  for fitting these to the observations were computed with  {\sc lte-codes}, a package which includes {\sc sterne} \citep{behara06}, {\sc spectrum} \citep{jeffery01a}, {\sc lte\_lines} \citep{jeffery91} and {\sc sfit} \citep{jeffery01b}. These assume that the atmosphere is semi-infinite, plane parallel, and in radiative, hydrostatic, and local thermodynamic equilibrium (LTE).
    
    The first stage of analysis is to generate a grid of model atmospheres with {\sc sterne} and creating a model grid with three varying parameters: effective temperature (\teff), surface gravity ($g$), and helium fraction (by number: $n_{\rm He}$), and assuming solar abundances for other elements\footnote{In its simplest mode, {\sc sterne} requires abundances for hydrogen, helium, carbon, nitrogen, oxygen, silicon, calcium and iron. The abundances of other elements are scaled to the abundances for calcium and iron.}. These models are used as input to {\sc spectrum} to generate high resolution model spectra. {\sc sfit} compares these to the observed spectrum to obtain a first estimate for \teff, $g$ and $n_{\rm He}$. Based on these estimates, {\sc sfit} is used to make a first estimate of abundances by adjusting the abundances of the most significant metals, namely C, N, O, Si, Ca and Fe. Based on these, a new model atmosphere grid is generated with {\sc sterne} and the process is iterated.  A final {\sc sterne} model is computed with \teff, $g$, $n_{\rm He}$ and major abundances based on these results, and the remaining metal abundances are  measured from the observed spectrum using this model. The final composition is shown in Tables \ref{table:pg_abun}\ and \ref{table:uvo_abun}.
    
	Evidence that the LTE approximation is violated sometimes arises  when fitting the hydrogen Balmer lines (the Balmer problem). This is where lower order Balmer lines appear too weak, while higher order lines are too strong, resulting in a range of \teff\ fitting different regions of the spectrum \citep[cf.][]{latour15}. For this analysis the adopted value of \teff\ was the mean obtained after fitting five Balmer lines individually. While H$\alpha$ is not well fit by our models, there is good agreement between the other four lines, so that \teff, $\log{g}$, and helium fraction are obtained with confidence. 
    
	{\sc sfit} provides several algorithms designed to fit a model spectrum to the observed data. That used here is the Levenburg-Marquardt downhill simplex, which takes initial parameters and calculates the derivative of each with respect to $\chi^2_{\rm red}$. The algorithm then takes a step downhill by a distance determined by the steepness of the slope, and repeats this until a minimum $\chi^2_{\rm red}$ is found. As the algorithm approaches a minimum, finer steps are taken to improve accuracy. Once \teff, $g$, and $n_{\rm He}$ have been computed, abundances for each element or ion with atomic data available were fitted individually. Abundances for each element are measured separately for each spectral region observed, and then combined using a weighted mean. Errors given on the weighted mean are found by summing the reciprocal errors in quadrature.

	The Levenburg-Marquardt algorithm provides standard errors in its analysis of any parameter and these are generally included. However, for many areas of the spectra where line blanketing is too severe and the spectrum rarely reaches the continuum, the automated process may not distinguish between spectral lines and an extremely low signal to noise ratio. It is therefore unable to accurately fit the abundances, and may fit a straight line through the middle of the spectral lines. In order to proceed, the fit was conducted by hand and the abundance was manually adjusted until a best fit was identified. The errors were then determined by adjusting the abundance until the fit was judged to be unsatisfactory in a "$\chi$-by-eye" method. Standard errors are given where available. The number of absorption lines of each element having a theoretical equivalent width $W_{\lambda} > 5$m\AA\ that contribute to the fit in each spectral region is also recorded. 

In order to represent the quantity of any given element in each star, we employ the following formula for chemical abundance of element \textit{i}:
\begin{equation}
	\log \epsilon_i = \log \frac{n_i}{\Sigma_i n_i} + c
\end{equation}
where $c$ is determined such that $\log \Sigma_i \mu_i \epsilon_i + c = \log \Sigma_i \mu_i \epsilon_{i\odot} + c' = 12.15$ and $\mu_i$ is the atomic mass of element $i$. Abundances relative to solar values as given by \citet{asplund05} are given in logarithmic form and denoted by the atomic symbol in square brackets (e.g. $[{\rm C}] = \log (\epsilon_{\rm C}/\epsilon_{\rm C \odot}$).

\subsection{Atomic data}   
    The atomic data used in this analysis were compiled from several sources. We began with the database provided by \cite{uky17}. 
    This was augmented to include data on Zn, Ga, Ge, Sr, and Ba from \cite{TOSS15}. Where duplicate entries were found, the latter source was preferred due to being specific to those elements. The resulting data were then merged with the Kurucz atomic database \citep{kurucz16}. Again, where duplicate entries were found, the latter were used.
    
    We searched for Pb {\sc iv} lines using theoretical oscillator strengths from \cite{alonso11}. These data were extracted and merged into our atomic line database.

	Because both stars have temperatures in the region of 40\,000K, there are substantial fractions of highly ionised species. Due to the conditions that produce these ions being difficult to recreate,  experimental oscillator strengths are frequently either poor or missing. Theoretical values exist for some ions, and were used for the heavier elements where no experimental data are available. 
    
    Despite gathering as much data as possible, both stars contain a large fraction of lines that we were unable to model. For example, when analysing the FUV region of \pgstar, a total of 806 lines, roughly 20\% of an estimated 3,600 lines, were successfully used to extract abundances. Similarly, in the NUV, an estimated 2,400 lines are detected, and of these, 936 (39\%), were successfully identified and contributed to abundance measurements. 

\begin{figure}
\centering
\includegraphics[scale = 1]{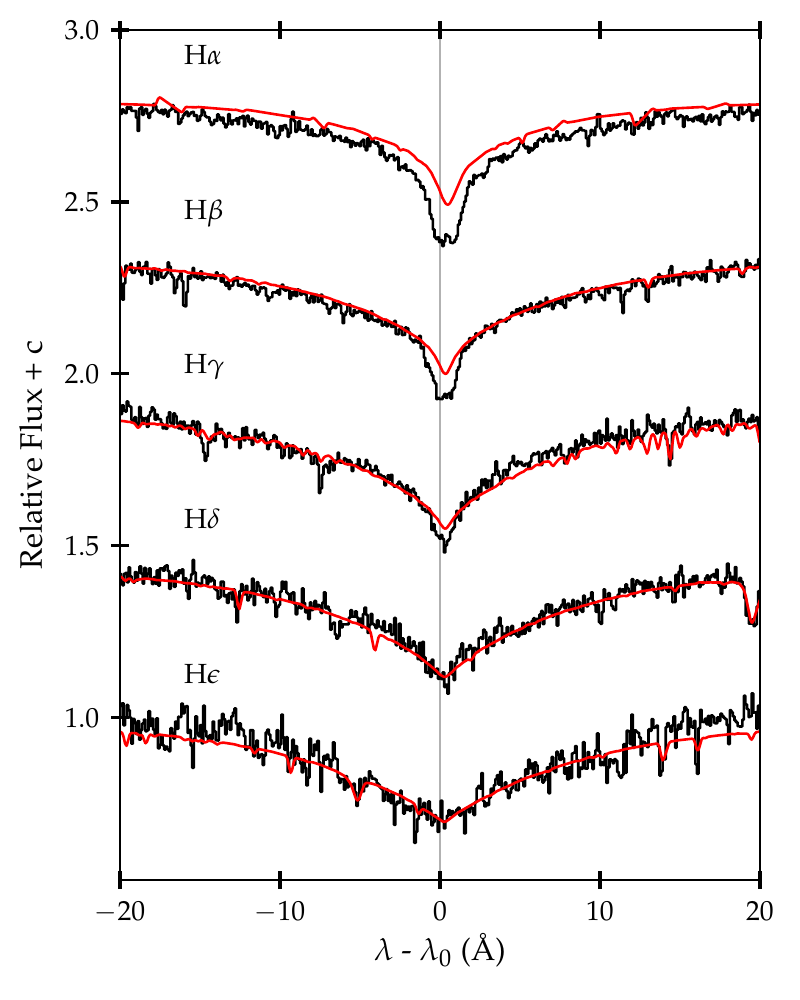}
\caption{ The Balmer lines of \pgstar, showing the difficulty in fitting the hydrogen line cores and wings simultaneously.}
\label{fig:pg_balmer}
\end{figure}

\begin{figure}
\centering
\includegraphics[scale = 1.0]{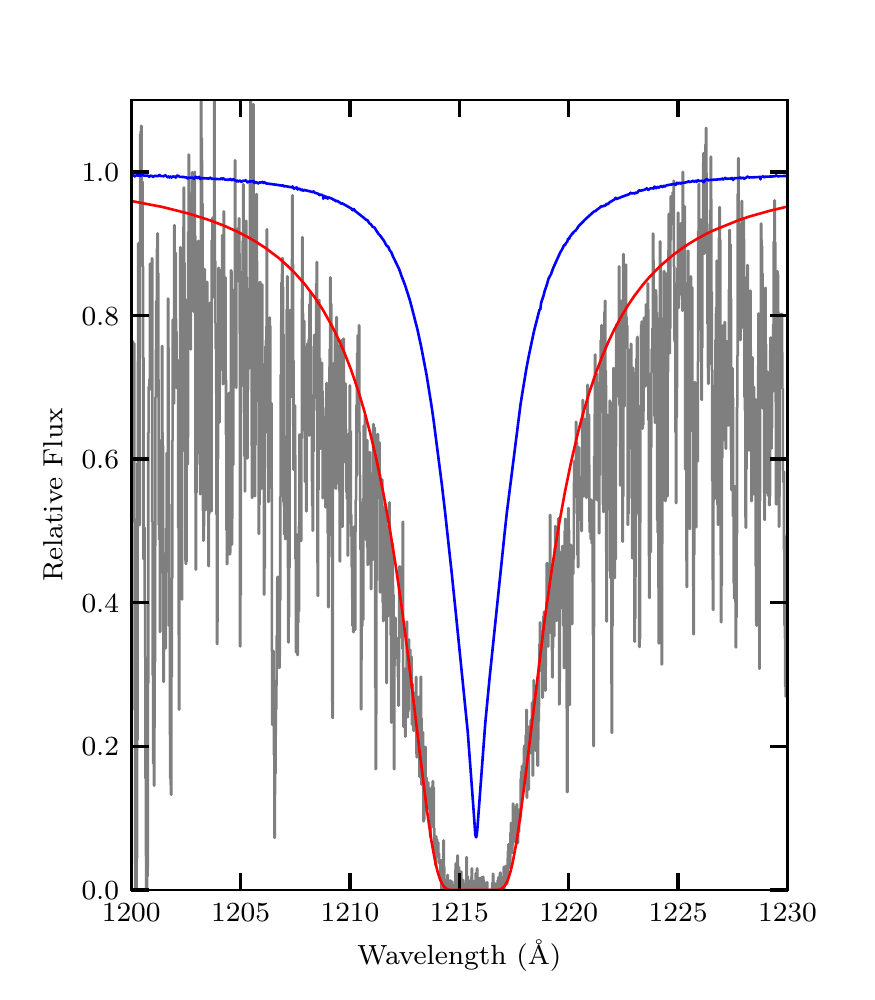}
\caption{ The FUV spectrum of \pgstar\ in the region of Ly$\alpha$ (grey) is shown together with the theoretical photospheric profile assuming $\nHe = 0.126$ (blue) and the theoretical interstellar profile for a hydrogen column density of $2.39 \times 10^{+20} {\rm cm^{-2}}$ (red).}
\label{fig:pg_lymanA}
\end{figure}

\begin{figure}
\centering
\includegraphics{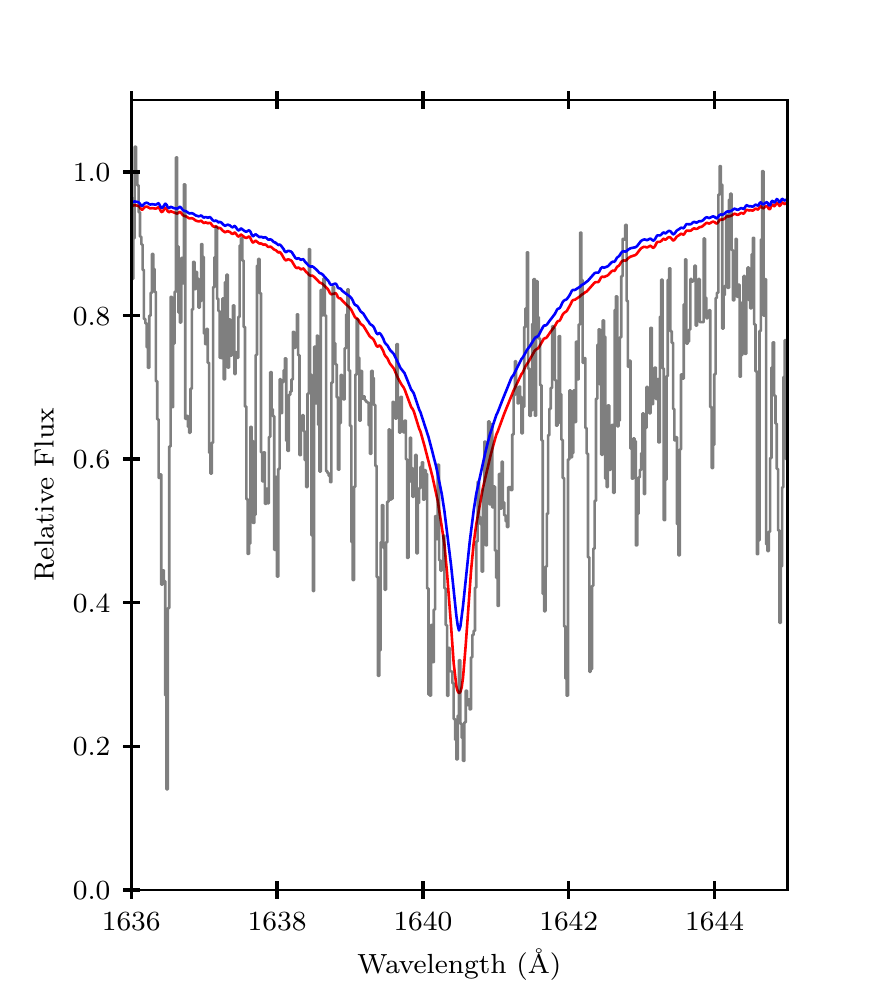}
\caption{The observed spectrum of \pgstar\ around the He {\sc ii} 1640.37\AA\ line (grey). The theoretical line is modelled as a singlet (red) and including fine structure (blue).}
\label{fig:pg_he1640}
\end{figure}

\section{Results: PG 0909+276}
\label{section:pg0909}
\subsection{Basic Parameters}

\begin{table}
\centering
\setlength{\tabcolsep}{2pt}
\begin{tabular}{cccccc}
\toprule
Star / Data		& \teff/K			& \lgcs			& \nHe			& E(B-V)\\
\toprule
\multicolumn{2}{l}{\pgstar} 	 \\
 ~~DSAZ		& 37,290$\pm$640	& 6.10$\pm$0.20	& 0.126$\pm$0.008	& 0.05	\\
 ~~IUE		& 37,160$\pm$220	& -				& -					& -		\\[2mm]
\multicolumn{2}{l}{\uvstar} 	\\
 ~~DSAZ		& 36,670$\pm$1,430	& 5.75$\pm$0.16	& 0.159$\pm$0.062	& 0.05	\\
 ~~IUE		& 36,670$\pm$260	& -				& - 				& -		\\
\toprule
\end{tabular}
\caption{Basic parameters derived for programme stars from high-resolution spectroscopy (STIS) and spectrophotometry (IUE).}
\label{table:basic_params}
\end{table}

\begin{table*}
\centering
\begin{tabular}{c ccc ccc ccc}
\toprule
Z & N & Mean & $\log \epsilon/\epsilon_{\odot}$ & FUV & NUV & OP & NUV-FUV & OP-FUV & OP-NUV \\
\toprule

H & 0/0/5 & $11.70\pm0.10$ & -0.30 & $-$ & $-$ & $11.70\pm0.10$ & $-$ & $-$ & $-$ \\
He & 1/0/14 & $11.04\pm0.12$ & 0.11 & $11.05\pm0.15$ & $-$ & $11.02\pm0.18$ & $-$ & $-0.03\pm0.23$ & $-$ \\
C & 30/12/32 & $8.68\pm0.12$ & 0.25 & $8.62\pm0.18$ & $8.78\pm0.20$ & $8.61\pm0.30$ & $0.16\pm0.27$ & $-0.01\pm0.35$ & $-0.17\pm0.36$ \\
N & 2/5/15 & $7.93\pm0.13$ & 0.10 & $7.80\pm0.48$ & $7.77\pm0.25$ & $8.02\pm0.17$ & $-0.03\pm0.54$ & $0.22\pm0.51$ & $0.25\pm0.30$ \\
Si & 9/0/0 & $5.95\pm0.20$ & -1.56 & $5.95\pm0.20$ & $-$ & $-$ & $-$ & $-$ & $-$ \\
S & 36/36/8 & $8.20\pm0.14$ & 1.08 & $8.59\pm0.25$ & $8.50\pm0.30$ & $7.81\pm0.20$ & $-0.09\pm0.39$ & $-0.78\pm0.32$ & $-0.69\pm0.36$ \\
Ar & 34/35/2 & $8.36\pm0.11$ & 1.96 & $8.29\pm0.28$ & $8.41\pm0.15$ & $8.30\pm0.20$ & $0.12\pm0.32$ & $0.01\pm0.34$ & $-0.11\pm0.25$ \\
Ca & 25/30/15 & $8.20\pm0.10$ & 1.86 & $8.30\pm0.27$ & $8.37\pm0.13$ & $7.73\pm0.20$ & $0.07\pm0.30$ & $-0.57\pm0.34$ & $-0.64\pm0.24$ \\
Sc & 9/28/39 & $7.66\pm0.12$ & 4.51 & $7.64\pm0.25$ & $7.30\pm0.30$ & $7.75\pm0.15$ & $-0.34\pm0.39$ & $0.11\pm0.29$ & $0.45\pm0.34$ \\
Ti & 0/14/24 & $7.69\pm0.11$ & 2.74 & $-$ & $7.50\pm0.23$ & $7.75\pm0.13$ & $-$ & $-$ & $0.25\pm0.26$ \\
V & 6/58/5 & $7.32\pm0.13$ & 3.39 & $6.30\pm0.25$ & $7.49\pm0.18$ & $8.00\pm0.25$ & $1.19\pm0.31$ & $1.70\pm0.35$ & $0.51\pm0.31$ \\
Cr & 79/150/0 & $7.59\pm0.17$ & 1.95 & $7.45\pm0.23$ & $7.75\pm0.25$ & $-$ & $0.30\pm0.34$ & $-$ & $-$ \\
Mn & 0/46/0 & $7.02\pm0.15$ & 1.59 & $-$ & $7.02\pm0.15$ & $-$ & $-$ & $-$ & $-$ \\
Fe & 67/0/0 & $7.10\pm0.20$ & -0.40 & $7.10\pm0.20$ & $-$ & $-$ & $-$ & $-$ & $-$ \\
Co & 160/328/0 & $7.97\pm0.17$ & 2.98 & $7.73\pm0.23$ & $8.25\pm0.25$ & $-$ & $0.52\pm0.34$ & $-$ & $-$ \\
Ni & 277/186/0 & $7.88\pm0.14$ & 1.66 & $7.85\pm0.15$ & $8.00\pm0.33$ & $-$ & $0.15\pm0.36$ & $-$ & $-$ \\
Cu & 56/7/0 & $7.18\pm0.15$ & 2.99 & $6.90\pm0.28$ & $7.30\pm0.18$ & $-$ & $0.40\pm0.33$ & $-$ & $-$ \\
Zn & 14/0/0 & $6.50\pm0.15$ & 1.94 & $6.50\pm0.15$ & $-$ & $-$ & $-$ & $-$ & $-$ \\
Ga & 14/0/0 & $4.90\pm0.15$ & 1.86 & $4.90\pm0.15$ & $-$ & $-$ & $-$ & $-$ & $-$ \\
Ge & 2/0/0 & $6.80\pm0.20$ & 3.15 & $6.80\pm0.20$ & $-$ & $-$ & $-$ & $-$ & $-$ \\
Sn & 1/0/0 & $3.60\pm0.20$ & 1.56 & $3.60\pm0.20$ & $-$ & $-$ & $-$ & $-$ & $-$ \\
Pb & 1/1/0 & $4.29\pm0.12$ & 2.54 & $4.40\pm0.15$ & $4.10\pm0.20$ & $-$ & $-0.30\pm0.25$ & $-$ & $-$ \\
\toprule
\end{tabular}
\caption{Abundance results for \pgstar\ in the three spectral regions, and the mean weighted by the square of the errors. The differences, where available, between abundance measurements are shown. The standard deviation is presented here not as a statistical tool, but rather as a loose indicator due to only using two or three data.}
\label{table:pg_abun}
\end{table*}

The three wavelength regions were treated independently. 
The optical region (3900\AA - 6900\AA) contains the majority of  hydrogen and helium lines, which  were used to fit the basic parameters: $T_{\rm eff}$, $g$ and H and He number fractions. Since these parameters are strongly correlated, they were fitted simultaneously. The best fit with formal errors is $\teff =  37,290\pm640$K, $\lgcs = 6.10\pm0.20$, and helium number fraction  $\nHe = 0.126\pm0.008$.

\pgstar\ expressed the Balmer problem, shown in Fig. \ref{fig:pg_balmer}, with the H$\alpha$ line showing the largest deviation. With good agreement between the other four lines, we have confidence in the adopted  parameters. 

	The FUV Lyman $\alpha$ line could not be modelled by stellar absorption in either star. In both cases, the line is successfully modelled by including absorption due to interstellar hydrogen \cite{lamers89}. This feature was best fit by a model with hydrogen column density of $2.39 \times 10^{+20} cm^{-2}$ (Fig. \ref{fig:pg_lymanA})

	We had difficulty fitting He {\sc ii} 1640.37\AA. Unexpected  structure in the core was poorly fitted when the line was modelled as a single line \citep{schoning89}. He {\sc ii} 1640.37\AA\ consists of a multiplet with seven fine-structure components. We improved the theoretical profile by treating all seven components individually, using their respective wavelengths and statistical weights from LS coupling. The result did not reproduce the observed structure, instead we observed a slightly weakened overall line strength (Fig \ref{fig:pg_he1640}). The observed structure is likely due to a combination of non-LTE processes and unidentified lines from  other atomic species.
    
\subsection{Non-Stellar Lines}
\begin{table}
\centering
\begin{tabular}{c c c c c c}
\toprule
4403.42 & 4403.71 & 4485.71 & 4504.63 & 4435.25 & 4440.35 \\
4485.71 & 4488.68 & 4504.26 & 4916.12 & 4948.97 & 4964.12 \\
4968.75 & 4976.59 & 4992.13 & 5070.78 & 5106.67 & 5219.11 \\
5243.74 & 5352.99 & 5399.17 \\
\toprule
\end{tabular}
\caption{Wavelengths of lines in the spectrum of \pgstar\ attributed to non-stellar thorium, and excluded from fits.}
\label{table:pg_th_lines}
\end{table}

	Once the basic parameters had been determined, we identified as many non-stellar lines as possible. 
    We checked Table 2 of \cite{morton78} for known interstellar lines, and masked these from the data in subsequent analyses. 
    The optical spectrum of \pgstar\ also contained a number of strong neutral thorium  lines. These presumably arise from contamination by the wavelength calibration lamp. They were identified and masked in the same fashion as interstellar lines. A list is included in Table \ref{table:pg_th_lines}. In our analysis of \uvstar, we used the same procedure to mask interstellar features, and did not detect neutral thorium lines.

\subsection{C, N, O, Si, Ca, Fe}
	As the model atmosphere requires some metal abundances as an input, these were the first to be measured. The final measured values for all elements are contained in Table \ref{table:pg_abun} with errors, including the weighted mean.
    
    Carbon was found to have a mean abundance $\log \epsilon_{\rm C}=8.68\pm0.12$ over the three regions, with very good agreement. A good number of lines were present in all three regions. 
    The nitrogen abundance is well-constrained across all three data sets with a mean value $\log \epsilon_{\rm N} = 7.93\pm0.13$. 
    We were unable to find an abundance for oxygen, as there were no observed lines in the spectrum. We found a limit $\log \epsilon_{\rm O}\lesssim 6.00$ above which O {\sc iv} 1343.51\AA\ should be observable. 
    The carbon and nitrogen abundances are close to solar, in agreement with \citet{edelmann03th}. 
	
	Obtaining an abundance for iron is limited by the lack of clear iron lines in the optical and NUV. In both regions we constrain $\log \epsilon_{\rm Fe}\lesssim 7.50$; at this abundance unobserved lines begin to exceed the noise. The FUV contains several iron  lines; and these give an abundance $\log \epsilon_{\rm Fe} = 7.10\pm0.20$. This comes from a sample of lines of which roughly half are blended with another element. Several strong unobserved lines begin to appear above abundances of $\log \epsilon_{\rm Fe} =  7.30$, which we adopt as the upper limit.

	Silicon is poorly represented with only 9 lines detected throughout the whole spectrum, all in the FUV. These lines are in good agreement with each other and give $\log \epsilon_{\rm Si}=5.95\pm0.20$. Contrary to C, N and O, the silicon abundance is substantially lower than solar at $\rm [Si] = -1.560\pm0.200$. 
    
    Calcium shows 70 lines across the spectra. As the FUV and NUV abundances agree to within errors and the optical abundance does not, we combined the UV measurements to obtain a mean abundance $\log \epsilon_{\rm Ca} = 8.55\pm0.19$, compared to $\log \epsilon_{\rm Ca} = 7.73\pm0.20$ in the optical. This gives a discrepancy of $0.82\pm0.28$ between UV and optical, the implications of which are discussed in section \ref{section:discussion}

\subsection{Inconsistent Abundance Measurements}

	Table \ref{table:pg_abun} shows the differences between observed abundances for elements measured in  more than one region. While many elements were either seen in only one region, or had abundances that were in agreement, we observed a $2\sigma$\ discrepancy in three elements: S, Ca, and V.

	We observed substantial depletion in the optical relative to the UV in both sulphur and calcium. However, we observed agreement across the UV in both. Thus, we take the mean of the FUV and NUV as one value and the optical as another value. For sulphur, we then have a UV abundance of $\log \epsilon_{\rm S} = 8.55\pm0.19$. We calculated a discrepancy between the optical and UV of $-0.74\pm0.27$. Similarly, calcium has a UV abundance of $\log \epsilon_{\rm Ca} = 8.35\pm0.12$. We also find a mean depletion in the optical of $0.63\pm0.23$.
    
    Vanadium was observed to have a significant disagreement between two regions, with differences of $1.19\pm0.31$ between the NUV and FUV, and $1.70\pm0.35$ between the optical and FUV. 
        
\begin{figure}
\includegraphics[scale=1.0]{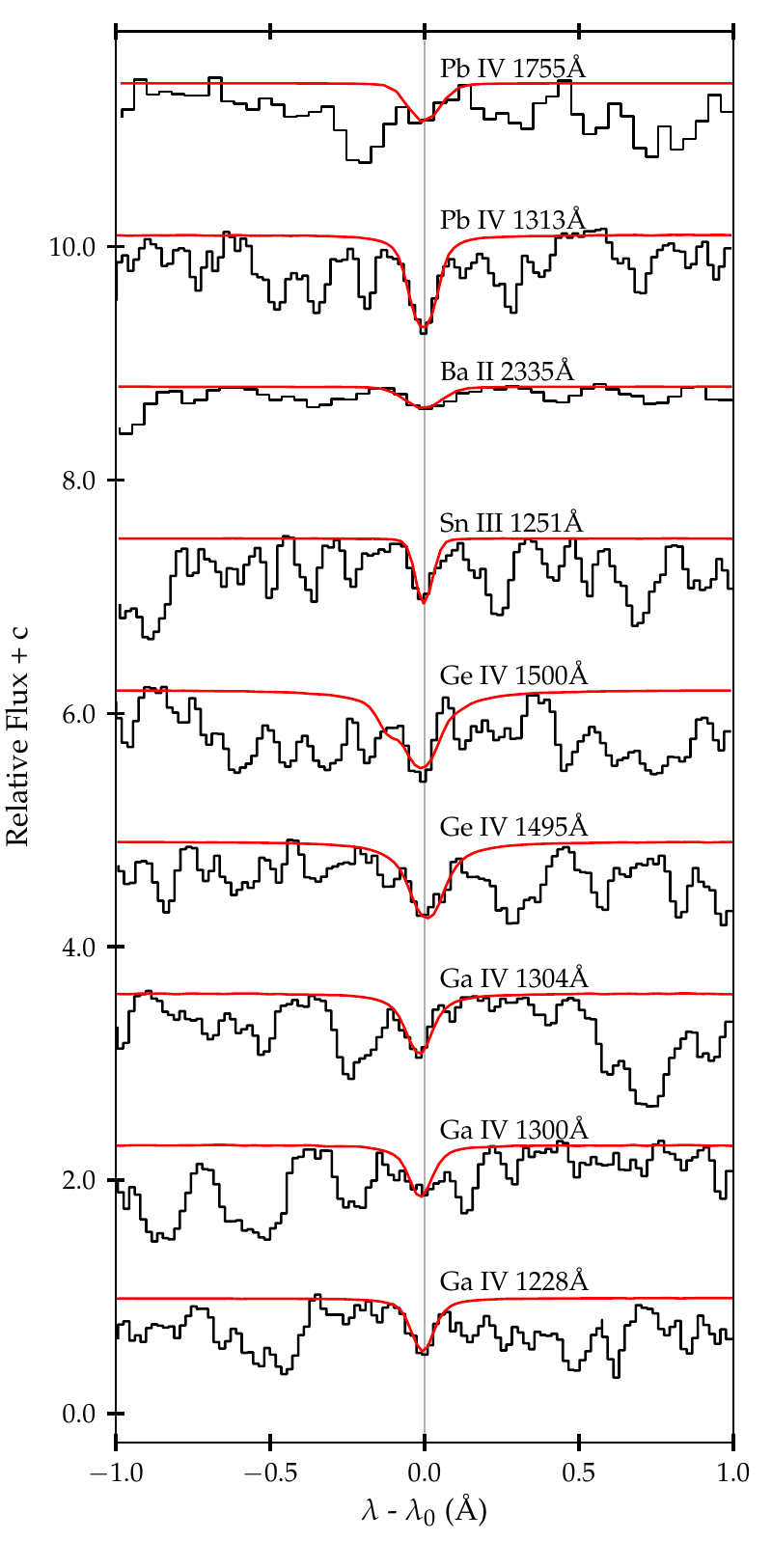}
\caption{Line profiles of barium, lead, tin, gallium, and germanium lines for \pgstar, showing STIS spectrum (black histogram) and best-fit model (red).}
\label{Pb_Sn_fits}
\end{figure}

\subsection{Other Elements}

    
	We detected two lead lines, at 1313.10\AA\ and 1755.00\AA\, which were modelled with atomic data from \cite{alonso11}.  From them, we infer that \pgstar\ is rich in lead with an abundance of $\log \epsilon_{\rm Pb}=4.29\pm0.12$, or a number fraction of order 100 times solar. 
    
	We observed a line at 2335.26\AA\ which could be attributed to Ba {\sc ii} and from which we would measure $\log \epsilon_{\rm Ba} = 8.75 \pm 0.15$. We searched  other regions of  spectrum for barium  lines (e.g. around  4130.65\AA\, 4554.03\AA, and 4934.08\AA), without success. Whilst a barium abundance of 8.75 is consistent with the optical lines not being seen, it represents an overabundance of $>6$ dex and relies on a single line. We have therefore not shown barium in any of the summary tables (e.g. Table \ref{table:pg_abun}).
    
    We probed for Ga, Ge, and Sn, as significant enrichments were observed previously by \cite{otoole06} in their analysis. We observed \pgstar\ to have 14 gallium lines in the FUV consistent with an abundance $\log \epsilon_{\rm Ga} = 4.90\pm0.15$. Similar to the stars studied by \cite{otoole06}, \pgstar\ is enriched by 1.86$\pm$0.17 dex relative to a solar abundance of 3.04$\pm$0.09.
    
    We observed the strong resonance Sn {\sc iii} 1251.39\AA\ line. Unlike the previous stars, it does not occur as a blend. This line is very sensitive to abundance changes, and we were able to fit it to a model of $\log \epsilon_{\rm Sn} = 3.60\pm0.20$. This is a relatively mild enrichment, only 1.46 dex relative to solar. The fit to this Sn {\sc iii} line, along with the Ga, Ge, Ba, and Pb lines from \pgstar are shown in figure \ref{Pb_Sn_fits}. 
    
    We noted two spectral lines at  wavelengths \cite{otoole06} identified to be germanium, 1494.89\AA\ and 1500.61\AA. To match the line depths, we required an abundance of $\log \epsilon_{\rm Ge} = 6.80\pm0.25$. Although extremely enriched (3.15 dex), similar levels have been found in other chemically peculiar sdB stars \citep{naslim11,naslim13}. We checked NIST for alternative candidate lines without success. 
    
\begin{figure*}
\includegraphics[scale=1]{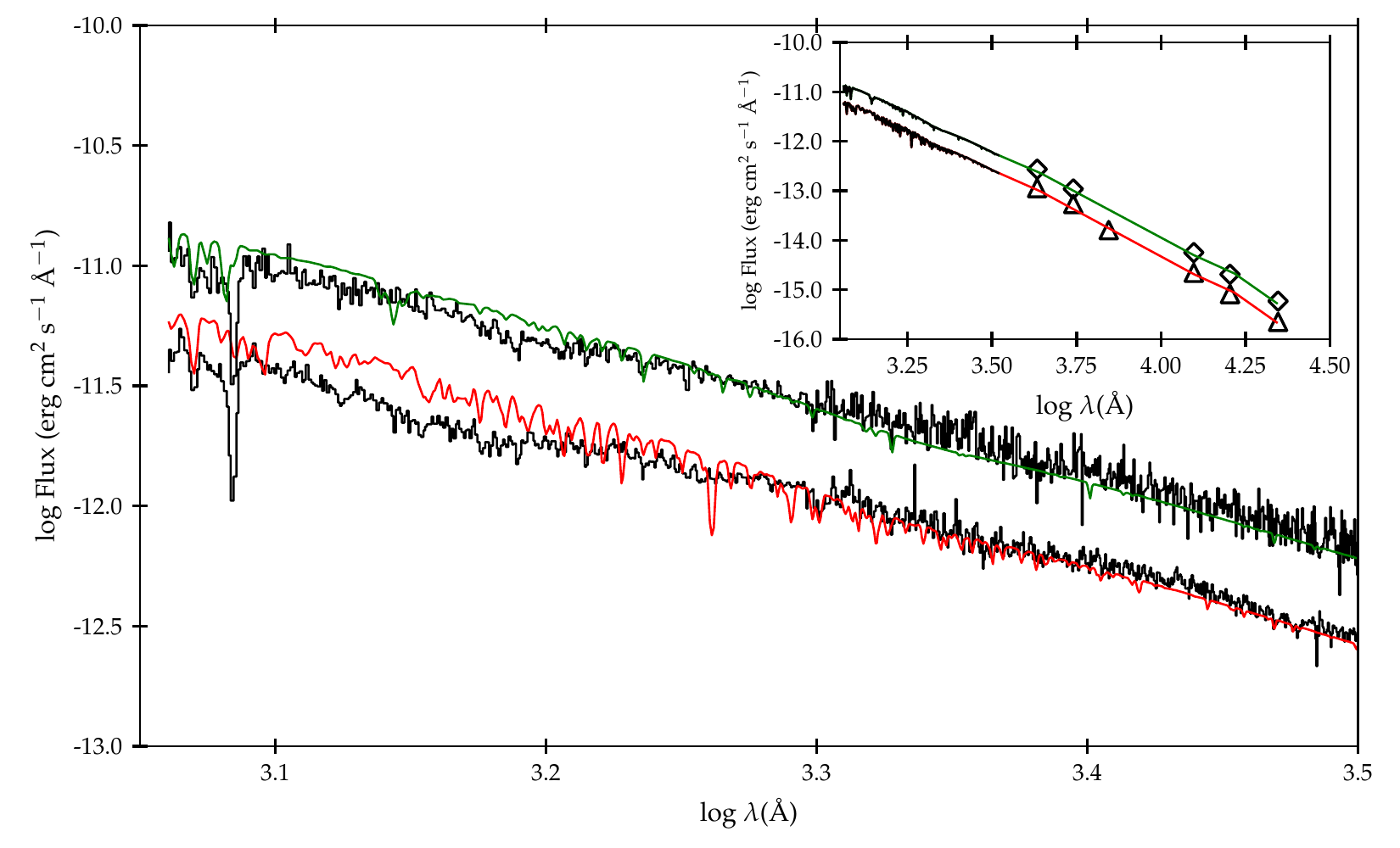}
\caption{The IUE low-resolution spectrophotometry for  \uvstar\ (upper black line) and \pgstar\ (lower black line).  The lower, red line is for a model with $\rm \teff = 37,160K$, and the green line is for a model with a $\rm \teff = 36,670$K. The inset compares models with broad-band photometry of \pgstar\ in  B, V, R, J, H, and K  \citep{hog00,zacharias09,cutri03} (triangles) and \uvstar\ in  B, V, J, H, and K \citep{hog00,cutri03} (diamonds).}
\label{fig:iue_pg}
\label{fig:iue_uvo}
\end{figure*}

\subsection{IUE Data}

	We compared the output flux of our final models with low-resolution IUE and optical and near infra-red photometry. Due to the degeneracy between \teff and \ebv, we fix \ebv\ at the hydrogen column density matching the Lyman $\alpha$ line. Figure ~\ref{fig:iue_pg} shows that for wavelengths $3.3 < \log{\lambda} < 3.5$, the models  match observation quite well, but the regions observed with STIS ($3.0 < \log{\lambda} < 3.3$) are  more poorly fitted.  Our best fit to the IUE data was for a model with $\teff = 37,290\pm220$, $\ebv = 0.05$, and angular radius $\theta = 8.78\pm0.01 \times 10^{-12}$ radians. These solutions also match the B, V, R, J, H, and K broad-band photometry well \citep{drilling13, hog00, zacharias09, cutri03}.  
    
    As already indicated, there is substantial line opacity missing from the models at FUV wavelengths. We investigated models in which (a) all metals were increased by factors of 0.5(0.5)3.0 dex and (b) individual metal abundances were matched to those obtained by fine analysis, where iron is not significantly enhanced.  Both tests assumed an homogeneous atmosphere. In case (a), we found substantial line blocking in the FUV region, but much of this appeared to come from a number of very strong lines which are not resolved in the IUE spectrum and for which there is no evidence in the HST/STI spectrum. In case (b), we found negligible difference compared with the models shown on Fig.~\ref{fig:iue_pg}.
    

\section{Results: UVO 0512--08}
\label{section:uvo0512}

\begin{table*}
\centering
\begin{tabular}{c ccc ccc ccc}
\toprule
Z & N & Mean & $\log \epsilon/\epsilon_{\odot}$ & FUV & NUV & OP & NUV-FUV & OP-FUV & OP-NUV \\
\toprule

H & 0/0/5 & $11.80\pm0.04$ & -0.20 & - & - & $11.80\pm0.04$ & - & - & - \\
He & 1/0/14 & $11.16\pm0.16$ & 0.23 & $11.20\pm0.29$ & - & $11.14\pm0.19$ & - & $-0.06\pm0.35$ & - \\
C & 22/40/49 & $8.54\pm0.10$ & 0.11 & $8.39\pm0.13$ & $9.03\pm0.22$ & $8.47\pm0.28$ & $0.64\pm0.26$ & $0.08\pm0.31$ & $-0.56\pm0.36$ \\
N & 7/13/13 & $7.67\pm0.16$ & -0.16 & $7.91\pm0.38$ & $7.68\pm0.22$ & $7.48\pm0.31$ & $-0.23\pm0.44$ & $-0.43\pm0.49$ & $-0.20\pm0.38$ \\
Ne & 0/13/9 & $8.32\pm0.25$ & 0.39 & - & $8.90\pm0.53$ & $8.15\pm0.29$ & - & - & $-0.75\pm0.60$ \\
Mg & 0/9/0 & $8.52\pm0.73$ & 0.92 & - & $8.52\pm0.73$ & - & - & - & - \\
Al & 1/0/0 & $5.50\pm0.58$ & -0.95 & $5.50\pm0.58$ & - & - & - & - & - \\
Si & 3/0/0 & $4.91\pm0.40$ & -2.60 & $4.91\pm0.40$ & - & - & - & - & - \\
S & 19/24/26 & $7.62\pm0.18$ & 0.50 & $7.19\pm0.27$ & $7.72\pm0.42$ & $8.14\pm0.31$ & $0.53\pm0.50$ & $0.95\pm0.41$ & $0.42\pm0.52$ \\
Ar & 55/67/14 & $8.39\pm0.13$ & 1.99 & $8.31\pm0.18$ & $8.57\pm0.23$ & $8.32\pm0.30$ & $0.26\pm0.29$ & $0.01\pm0.35$ & $-0.25\pm0.38$ \\
Ca & 59/82/34 & $8.51\pm0.14$ & 2.17 & $8.77\pm0.20$ & $8.55\pm0.26$ & $7.88\pm0.30$ & $-0.22\pm0.33$ & $-0.89\pm0.36$ & $-0.67\pm0.40$ \\
Sc & 2/2/7 & $7.20\pm0.23$ & 4.05 & $6.91\pm0.39$ & $6.68\pm1.36$ & $7.39\pm0.30$ & $-0.23\pm1.41$ & $0.48\pm0.49$ & $0.71\pm1.39$ \\
Ti & 17/34/9 & $7.99\pm0.19$ & 3.04 & $7.76\pm0.45$ & $7.76\pm0.30$ & $8.32\pm0.30$ & $0.00\pm0.54$ & $0.56\pm0.54$ & $0.56\pm0.42$ \\
V & 40/19/0 & $6.84\pm0.17$ & 2.91 & $6.72\pm0.20$ & $7.14\pm0.32$ & - & $0.42\pm0.38$ & - & - \\
Cr & 331/317/0 & $8.67\pm0.12$ & 3.03 & $8.20\pm0.70$ & $8.68\pm0.12$ & - & $0.48\pm0.71$ & - & - \\
Mn & 53/13/0 & $7.50\pm0.24$ & 2.07 & $7.52\pm0.34$ & $7.48\pm0.33$ & - & $-0.04\pm0.47$ & - & - \\
Fe & 348/342/15 & $8.53\pm0.04$ & 1.03 & $8.49\pm0.05$ & $8.70\pm0.09$ & $8.00\pm0.29$ & $0.21\pm0.10$ & $-0.49\pm0.29$ & $-0.70\pm0.30$ \\
Co & 221/354/0 & $8.45\pm0.09$ & 3.46 & $7.60\pm0.28$ & $8.56\pm0.10$ & - & $0.96\pm0.30$ & - & - \\
Ni & 119/117/0 & $7.25\pm0.08$ & 1.03 & $7.00\pm0.09$ & $8.26\pm0.18$ & - & $1.26\pm0.20$ & - & - \\
Cu & 53/6/0 & $6.74\pm0.12$ & 2.55 & $6.35\pm0.15$ & $7.30\pm0.18$ & - & $0.95\pm0.23$ & - & - \\
Zn & 101/4/0 & $7.25\pm0.16$ & 2.69 & $7.25\pm0.16$ & $7.30\pm2.23$ & - & $0.05\pm2.24$ & - & - \\
Ga & 8/0/0 & $4.65\pm0.15$ & 1.61 & $4.65\pm0.15$ & - & - & - & - & - \\
Ge & 2/0/0 & $6.60\pm0.20$ & 2.95 & $6.60\pm0.20$ & - & - & - & - & - \\
Sn & 1/0/0 & $4.40\pm0.20$ & 2.36 & $4.40\pm0.20$ & - & - & - & - & - \\
Pb & 1/1/0 & $4.22\pm0.13$ & 2.47 & $4.95\pm0.30$ & $4.04\pm0.15$ & - & $-0.91\pm0.34$ & - & - \\

\toprule
\end{tabular}
\caption{Abundance results for \uvstar\ in the three spectral regions, and the mean weighted by the square of the errors. The differences, where available, between abundance measurements are shown. The standard deviation is presented here not as a statistical tool, but rather as a loose indicator due to only using two or three data.}
\label{table:uvo_abun}
\end{table*}

\begin{figure}
\centering
\includegraphics[scale = 1]{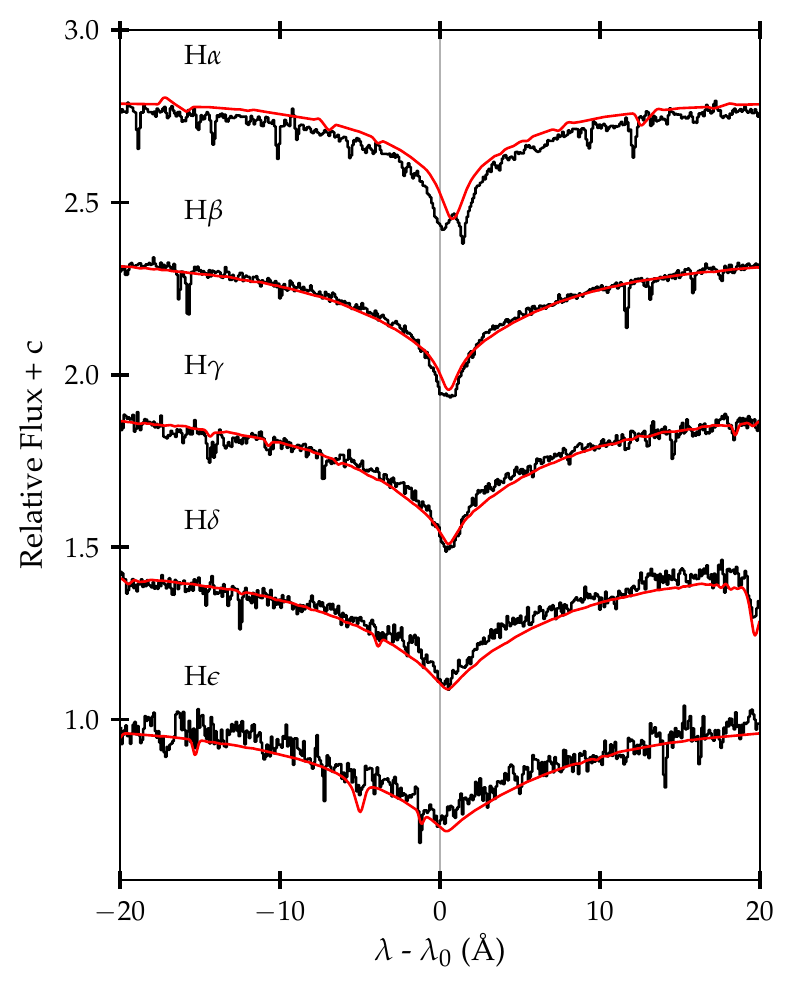}\\
\caption{The observed Balmer lines (black) and models (red) in \uvstar.}
\label{fig:uvo_balmer}
\end{figure}

\subsection{Basic Parameters}
\begin{figure}
\centering
\includegraphics[scale = 1]{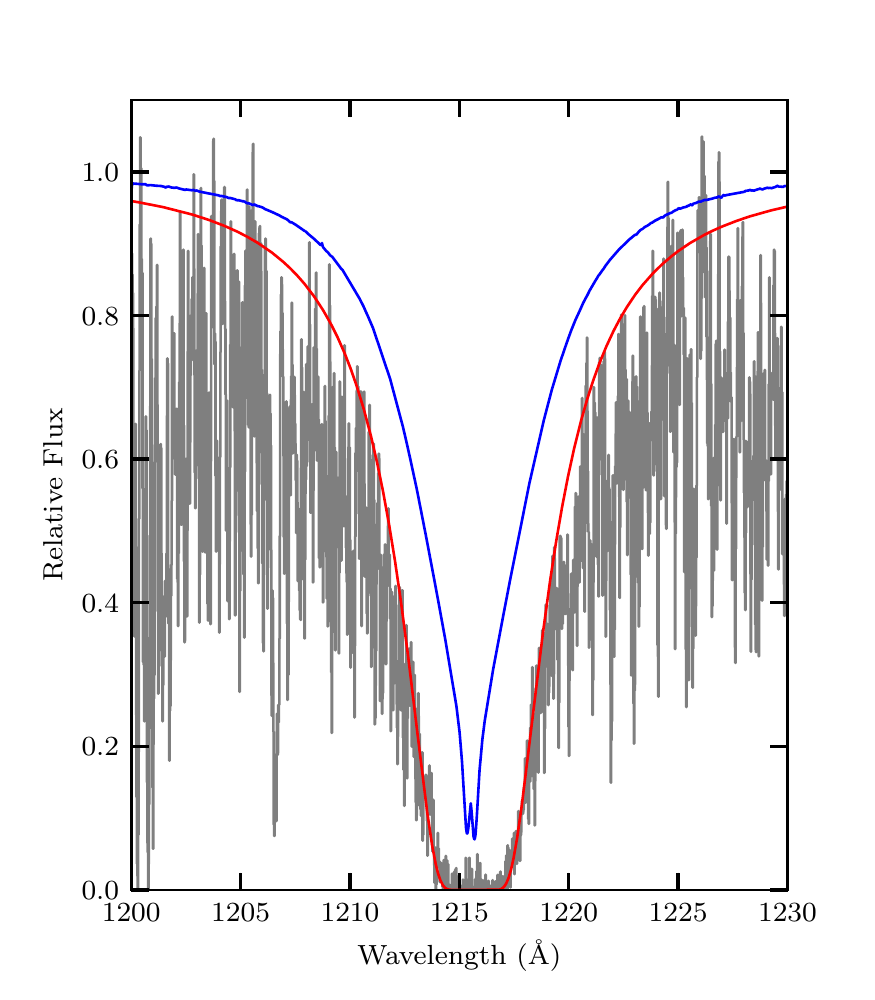}\\
\caption{The FUV spectrum of \uvstar\ in the region of Ly$\alpha$ (black) together with the theoretical photospheric profile assuming $\nHe = 0.159$ (blue) and the theoretical interstellar profile assuming a hydrogen column density of $\rm 2.39\times 10^{+20}cm^2$ } 
\label{fig:uvo_lymanA}
\end{figure}

	We performed the initial analysis for \uvstar\ identically to that for \pgstar. First, the optical H and He lines were fitted using a grid of atmosphere models in $\teff$, $\log g$ and helium fraction. This yielded values of $36,670\pm1,430$K, $5.75\pm0.16$, and $0.159\pm0.062$ respectively. Figure \ref{fig:uvo_balmer} shows the Balmer lines present in the optical region where it can be seen that the Balmer problem is more severe in \uvstar\ than in \pgstar. In particular,  H$\epsilon$ and H$\alpha$ lines are not well fitted. This is reflected in the larger errors attached to the $T_{\rm eff}$ measurement.

	Interstellar lines were removed in a similar fashion to \pgstar. We noted the same problem with modelling the 1640.37\AA\ He {\sc ii} line, and were similarly unable to reproduce the profile accurately. 

	The Lyman $\alpha$ line was also fitted to interstellar absorption, where we found a good fit to a hydrogen column density of $\rm 2.39\times10^{+20}cm^2$. Figure \ref{fig:uvo_lymanA} shows the difference between the best-fit stellar and interstellar models.

Measurements of individual elemental abundances for \uvstar\ were obtained (as for \pgstar) by optimizing the theoretical line spectrum to the observed spectra. Results for the three wavelength regions are given in Table~\ref{table:uvo_abun}. 

\subsection{C, N, O, Si, Ca, Fe}
\label{section:metals_uvo}

	We observed agreement in the carbon abundance between the FUV and optical about $\log \epsilon_{\rm C} = 8.48\pm0.07$;  in the NUV, we found $\log \epsilon_{\rm C} = 9.03\pm0.22$, a discrepancy of $0.55\pm0.25$ dex.

	We found only 19 nitrogen lines in the FUV and NUV, many of which were strong. We found a value close to  solar in all three regions, with a mean of ${\rm [N]} = -0.16$. 
    
    We did not observe any oxygen lines in any regions of the spectrum. In order for no lines with equivalent widths  $>$5m\AA\ to be seen, the abundance of oxygen in the star must be less than $\log \epsilon_{\rm O} \sim 5.50$, or ${\rm [O]} < -3.19$. Normal sdBs typically have $\rm [O] \approx -1$, making this a very oxygen-deficient subdwarf.

	For silicon, we only detected 3 strong lines in the FUV, giving an abundance ${\rm [Si]} = -2.60$. 
    In the case of calcium, the FUV and NUV abundances agreed well, yielding a weighted mean of $\log \epsilon_{\rm Ca} = 8.51\pm0.14$. The optical spectrum gave $\log \epsilon_{\rm Ca} = 7.88\pm0.30$, or a difference of $-0.63\pm0.33$. 

\begin{figure}
\centering
\includegraphics[scale=1]{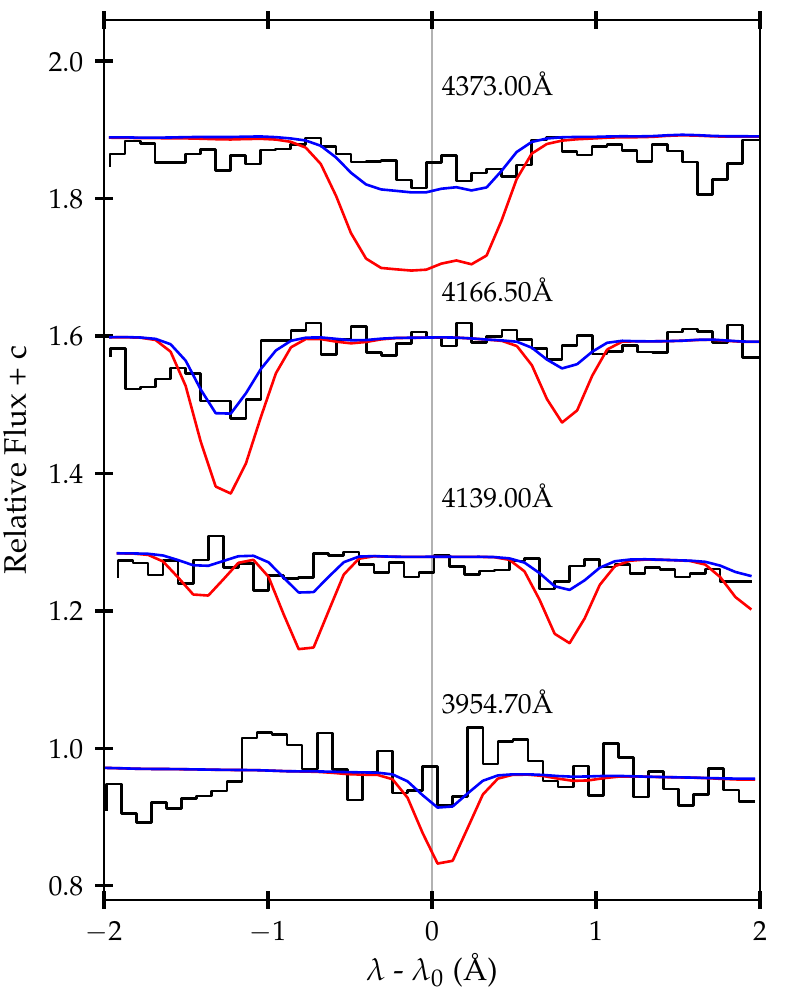}
\includegraphics[scale=1]{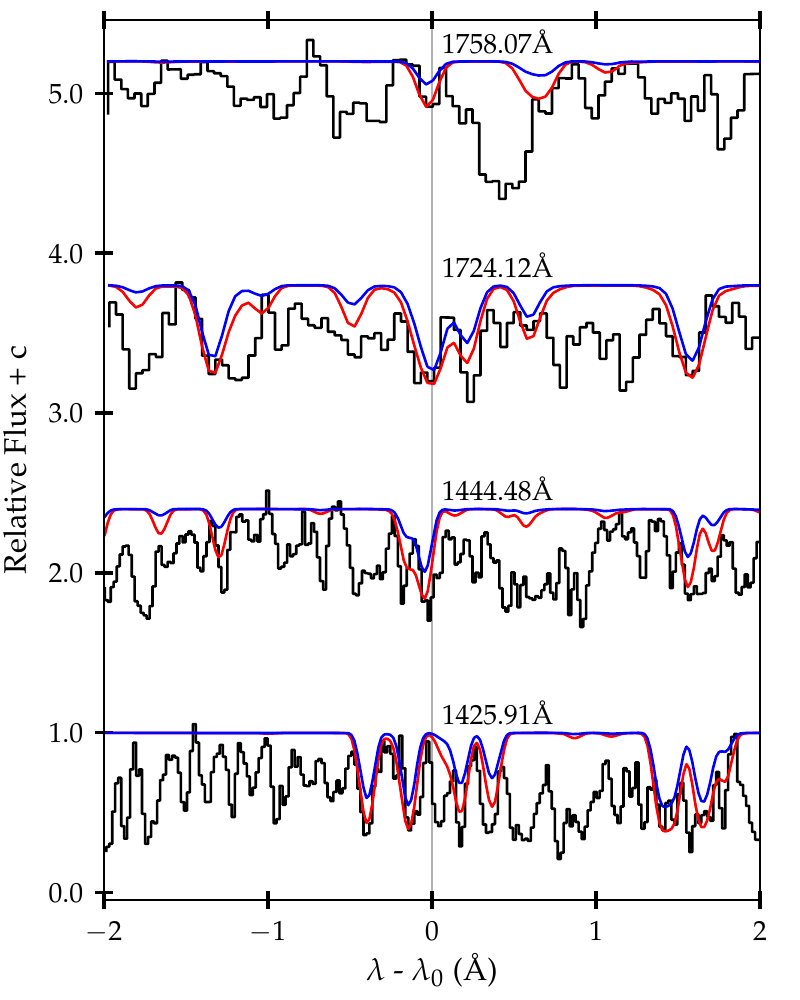}
\caption{Upper: Models of  Fe{\sc iii} lines in the optical region of \uvstar\ (black)  with $\log \epsilon_{\rm Fe} = 8.0$ (blue) and 8.7 (red). The labels above each observation denotes $\lambda_0$ for that section. Lower: The above plot is duplicated for the UV regions. These plots are separated due to differing scales.}
\label{fig:uvo_fe_comp}
\end{figure}

	We observed several hundred iron lines in the UV spectra of \uvstar, unlike \pgstar\ which expressed very few iron features. Agreement was observed between FUV and NUV spectra.  5 Weak Fe lines in the optical between 4318.20\AA\ and 4310.80\AA\ required an abundance lower by $0.57\pm0.21 $.  Fig. \ref{fig:uvo_fe_comp} demonstrates how lowering the optical abundance improves the fit to these lines.

\subsection{Inconsistent Abundance Measurements}
\label{section:uvo_strat}

For \uvstar, we detected abundance discrepancies exceeding $2\sigma$ between spectral regions in eight elements. Carbon, calcium and iron were discussed above. We also observed disagreement in sulphur, cobalt, nickel, copper and lead.

	Sulphur is well constrained in all three wavelength regions. FUV and NUV agree well, but the optical -- FUV difference is $0.95\pm0.41$. 
    
	Cobalt lines were seen in both FUV and NUV in very large numbers, with a large disagreement in abundance which is emphasized by the small error on the NUV measurement of $\log \epsilon_{\rm Co} = 8.56\pm0.10$. The NUV abundance is larger than the FUV value by $0.96\pm0.30$ dex. 

	We detected strongly differing quantities of nickel in the two UV regions, which both contain over 100 lines. The NUV abundance is larger by $1.26\pm0.20$ dex. 
    
    Copper shows an NUV abundance larger than the the FUV by $0.95\pm0.23$ dex. While more lines were detected in the FUV, they often occur in busier segments of the spectrum, requiring them to be fitted by hand. 
    
    The FUV and NUV titanium abundances are similar, but lower than the optical by $0.56\pm0.54$ and $0.56\pm0.42$ dex, respectively.
    
	Finally, we found a disagreement between abundances given by the two lead lines (1755.00\AA and 1313.10\AA) of $-0.91\pm0.34$ dex. The average of the two abundances gives $\log \epsilon_{\rm Pb} = 4.22\pm0.13$. The model fits are shown in figure \ref{fig:uvo_Pb_Ga_Sn}.

\subsection{Other Elements}

	We conducted a search for features due elements with $Z > 30$ for which atomic data are available. We found no traces of the majority, germanium, arsenic, strontium, yttrium, tin, and barium. 
    
	In the cases of magnesium, aluminium and silicon, we detected spectral lines in only one region. We observed magnesium to be enriched relative to other sdBs with $\rm [Mg] = 0.92$, in contrast to the sub-solar values of all other sdBs in figure \ref{fig:abs}. Aluminium is consistent with previous measurements of sdBs.
    
\begin{figure}
\includegraphics[scale=1.0]{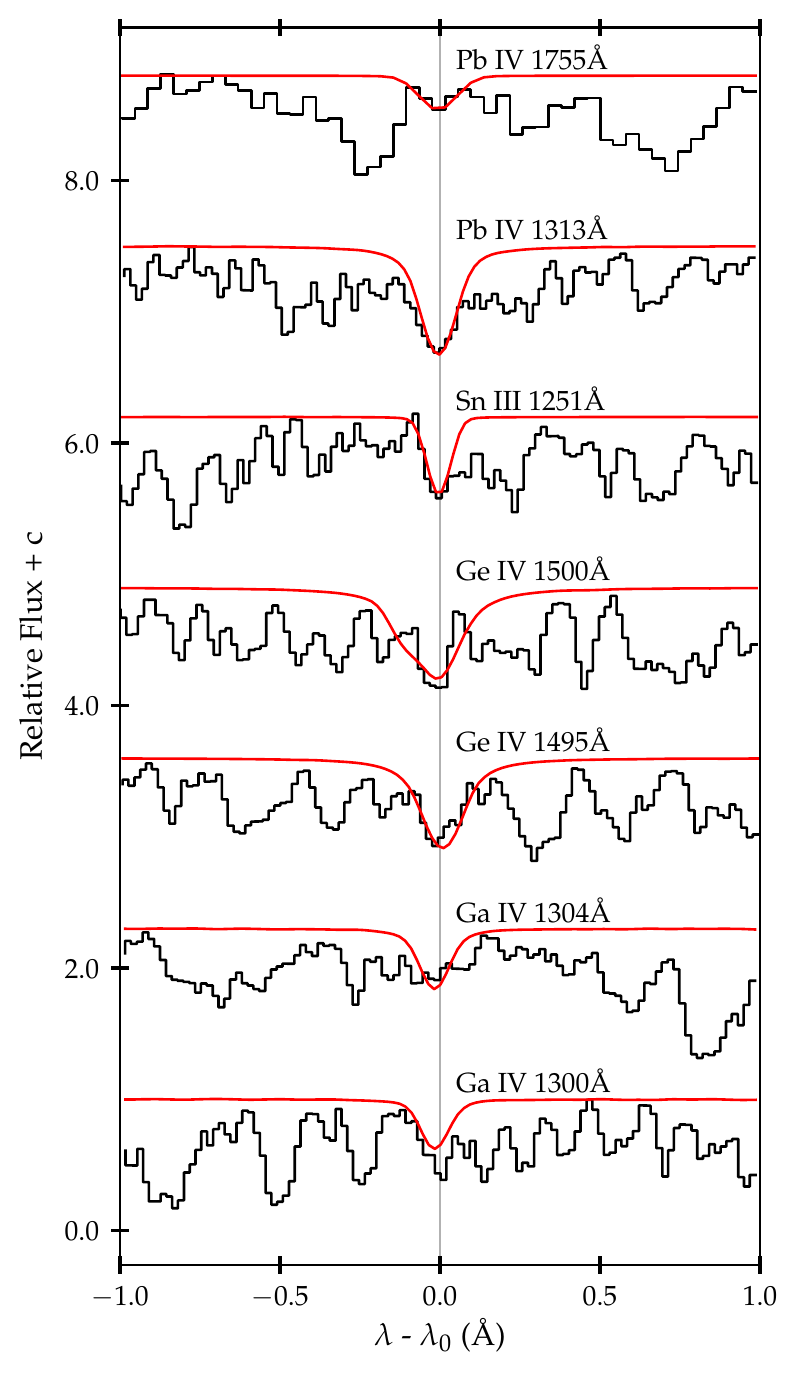}
\caption{A selection of \uvstar\ model Ga {\sc iv}, Ge{\sc iv}, Sn{\sc iii}, and Pb {\sc iv}\ lines (red). The STIS observations are shown in black.}
\label{fig:uvo_Pb_Ga_Sn}
\end{figure}

    While we did not observe any of the germanium lines seen in \pgstar, we detected eight gallium lines in the FUV, which gave us an abundance of $\log \epsilon_{\rm Ga} = 4.65\pm0.15$ (or [Ga]$ = 1.61\pm0.16$). 
    The Sn {\sc ii} 1251.39\AA\ line seen in \pgstar\ and by \cite{otoole06} was also seen in this star, and was well modelled by an enrichment of $\rm [Sn] = 2.36$. This is shown in figure \ref{fig:uvo_Pb_Ga_Sn}
    
	For a further six elements nitrogen, argon, scandium, chromium, manganese, and zinc, we detected no significant discrepancies across the wavelength regions in which they were observed, and no significant departures from the typical abundances of sdB stars.
    
    
\subsection{IUE Data}
	As in \pgstar, we fixed \ebv\ to match the column density of hydrogen found with the Lyman $\alpha$ line. Our fit to the IUE spectrophotometry for \uvstar\ gives $\teff = 36,670\pm260$K, $\ebv = 0.05$, and $\theta = 1.35\pm0.01 \times 10^{-11}$ radians (Fig. \ref{fig:iue_uvo}). \teff\ is consistent with the optical observations. The observed Ly$\alpha$ line can be seen at $\rm \log \lambda = 3.085$ and, as in the STIS spectroscopy, is dominated by interstellar hydrogen (cf. Fig.~\ref{fig:uvo_lymanA}). 

\begin{table*}
\centering
\caption{Elemental abundances for \pgstar, \uvstar, and related stars in the form  $\log \epsilon_i = \log n_i + c$ (see text). Measurement errors are shown in parenthesis. The absence of a reported measurement is indicated by ``$-$''.
}
\label{t:abs}
\setlength{\tabcolsep}{2pt}
\begin{tabular}{@{\extracolsep{0pt}}p{27mm}l lll lll lll lll}
\hline
Star 					& H			& He 		& C			& N			& O			& Ne		& Mg 		& Al		& Si		& S 		& Cl  		& Ar 		\\
\hline
PG\,0909$+$276$^{a}$	& 11.70(10)	& 11.04(12)	& 8.68(12)	& 7.94(13)	& $-$		& $-$		& $-$		& $-$		& 5.95(20)	& 8.20(14)	& $-$ 		& 8.36(11) 	\\
PG\,0909$+$276$^{1,2}$	&			& 11.15(10)	& 8.63(35)	& 8.00(23)	& $<6.00$	& $<7.87$	& $-$		& $<6.25$	& 5.80(10)	& 8.26(53)	& $-$ 		& 8.68(15) 	\\
UVO\,0512$-$08$^{a}$	& 11.80(04)	& 11.16(16)	& 8.54(10)	& 7.67(16)	& $-$		& 8.32(25)	& 8.52(73)	& 5.50(58)	& 4.91(40)	& 7.62(18)	& $-$ 		& 8.39(13) 	\\
UVO\,0512$-$08$^{1,2}$	&			& 11.23(10)	& 8.59(20)	& 7.94(21)	& $<5.50$	& $-$		& $-$		& $<6.25$	& 5.76		& 8.14(49)	& $-$ 		& 9.90 	  \\[1mm]
\ledz$^3$				& 11.8  	& 11.2	& $<6.5$		& 8.04(24)	& 7.43(07)	& 7.48(25)	&  6.25(11)	& 6.47(07)	&  6.26(21)	& 7.61(18)	& 6.34(11)	& $<8.3$		  \\
\crimson $^{4}$ 		& 11.83		& 11.23(05)	& 8.04(22)	& 8.02(20)	& 7.60(17)	& $<7.6$	& 6.85(10)	& $-$		& 6.32(12)	& $-$		& $-$ 		& $-$ 		 \\
HE\,1256$-$2738$^{5}$	& 11.45		& 11.44		& 8.90(54)	& 8.14(62)	& 8.08(10)	&$<7.1$		& $<6.5$	& $-$		& 6.19(10)	& $<6.5$	& $-$ 		& $-$  		 \\
HE\,2359$-$2844$^{5}$	& 11.58		& 11.38		& 8.51(29)	& 8.00(57)	& 7.81(16)	&$<6.9$		& 7.6(1)	& $-$		& 5.73(13)	& $<6.3$	& $-$ 		& $-$ 		 \\[1mm]
JL\,87$^{6}$			& 11.62(07)	& 11.26(18)	& 8.83(04)	& 8.77(23)	& 8.60(23)	& 8.31(57)	& 7.36(33)	& $-$		& 7.22(27)	&6.88(1.42)	& $-$		& $-$ 		  \\
cool sdB$^{2,b}$		&			& ~9.24(54)	& 6.99(47)	& 7.68(41)	& 7.88(26)	& $-$		& 6.54(26)	& 5.70(18)	& 6.79(37)	& 6.51(21)	& $-$ 		& 6.78(21)  \\
warm sdB$^{2,c}$		&			& 10.15(76)	& 7.73(70)	& 7.42(27)	& 7.67(51)	& 7.27(67)	& 7.17(29)	& 6.2		& 6.02(55)	& 7.18(56)	& $-$ 		& 7.89(17) 	\\
Feige\,66$^{7}$			&			& 10.4		& 6.79(30)	& 7.65(15)	& $-$		& $-$		& $-$ 		& $<3.5$	& $<2.0$	& 7.69(46)	& $-$ 		& 7.86(24) 	  \\[1mm]
Sun$^{8}$				& 12.00		& 10.93		& 8.43		& 7.83		& 8.69		& 7.93 		&  7.60		& 6.45		& 7.51		& 7.12		& 5.50		& 6.40		    \\[3mm]
\hline
Star					& Ca 		& Ti 		& V 		& Cr 		& Fe		& Co 		& Ni 		& Cu 		& Ga		& Ge		& Sn		& Pb \\
\hline
PG\,0909$+$276$^{a}$ 	& 8.20(10) 	& 7.69(13) 	& 7.32(13)	& 7.59(17)	& 7.10(20) 	& 7.97(17)	& 7.88(14)	& 7.18(15)	& 4.90(15)	& 6.80(20)	& 3.60(20)	& 4.29(12) \\
PG\,0909$+$276$^{1,2}$ 	& 7.81(35) 	&  7.97(20) & 8.10(26) 	& $-$ 		& $<7.87$ 	& $-$ 		& $-$ 		& $-$ 		& $-$		& $-$		& $-$		& $-$\\
UVO\,0512$-$08$^{a}$ 	& 8.51(14) 	& 7.99(19) 	& 6.84(17) 	& 8.67(12)	& 8.53(04) 	& 8.45(09)	& 7.25(08)	& 6.74(12)	& 4.65(15)	& 6.60(20)	& 4.40(20)	& 4.22(13) \\
UVO\,0512$-$08$^{1,2}$ 	&  8.10(24)	& 8.06(33) 	& 7.36(22)  & $-$ 		& $<7.81$ 	& $-$ 		& $-$ 		& $-$ 		& $-$		& $-$		& $-$		& $-$\\[1mm]
\ledz$^3$        		& 8.31(21) 	& 7.37(34)  & 7.51(25) 	& $-$ 		& $<7.0$ 	& $-$ 		& $-$ 		& $-$ 		& $-$		& 6.24(06)	& $-$		& 5.49(18) \\
\crimson $^{4}$ 		& $-$ 		& $<6.0$ 	& $<6.5$ 	& $<7.0$ 	& $<6.8$	& $-$ 		& $-$ 		& $-$ 		& $-$		& 6.28(12)	& $-$		& $-$\\
HE\,1256$-$2738$^{5}$ 	& $-$ 		& $-$ 		& $-$ 		& $-$ 		& $-$		& $-$ 		& $-$ 		& $-$ 		& $-$		& $-$		& $-$		& 6.39(23)\\
HE\,2359$-$2844$^{5}$ 	& $-$ 		& $-$ 		& $-$ 		& $-$ 		& $-$ 		& $-$ 		& $-$ 		& $-$ 		& $-$		& $-$		& $-$		& 5.64(16) \\[1mm]
JL\,87$^{6}$   			& $-$ 		& $-$ 		& $-$ 		& $-$ 		&&  $-$    	& $-$ 		& $-$ 		& $-$ 		& $-$		& $-$		& $-$		& $-$ \\
cool sdB$^{2,b}$		& $-$ 		& 6.30(35) 	& 7.10(36) 	& $-$ 		& 7.58(20) 	& $-$ 		& $-$ 		& $-$ 		& $-$		& $-$		& $-$		& $-$ \\
warm sdB$^{2,c}$    	& 7.98(25) 	& 7.04(36) 	& 7.78(20) 	& $-$ 		& 7.46(24) 	& $-$ 		& $-$ 		& $-$ 		& $-$		& $-$		& $-$		& $-$ \\
Feige\,66$^{7}$   		& 8.09(20)	& 6.96(22) 	& 6.37(22) 	& $-$		& 6.46(17) 	& $-$ 		& $-$ 		& $-$ 		& $-$		& 5.21(05)	& $-$		& 4.7  \\[1mm]
Sun$^{8}$  				& 6.34 		& 4.95 		& 3.93 		& 5.64 		& 7.50      & 4.99 		& 6.22 		& 4.19 		& 3.04		& 3.65		& 2.04		& 1.75 \\
\hline
\end{tabular}\\
\parbox{170mm}{
Notes:\\
$a$. This paper; abundances from spectral synthesis $\chi^2$ minimization. \\
$b$. $25\leq \teff/{\rm kK}\leq 27$. \\
$c$. $35\leq \teff/{\rm kK}\leq 40$ excluding PG\,0909$+$276 and UVO\,0512$-$08.\\
}\\[1mm]
\parbox{170mm}{
References: 
1. \citet{edelmann03},
2. \citet{geier13},
3. \citet{jeffery17a}
4. \citet{naslim11}, 
5. \citet{naslim13},
6. \citet{ahmad07},
7. \citet{otoole06}, 
8. \citet{asplund09}; photospheric except helium (helio-seismic), neon and argon (coronal). 
}
\end{table*}

\section{Discussion}
\label{section:discussion}

\begin{figure*}
\includegraphics[height=.42\textheight]{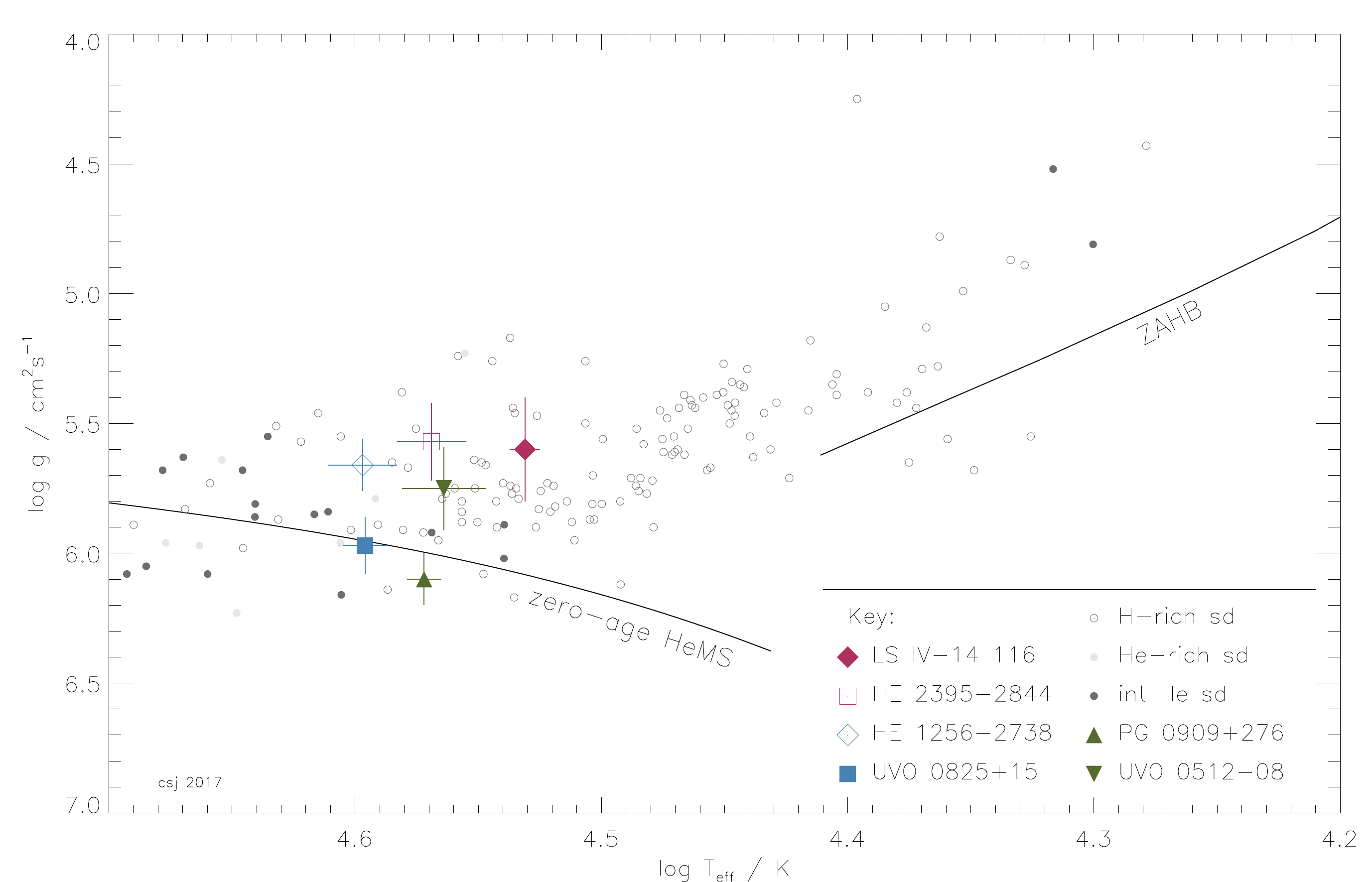}
\includegraphics[height=.42\textheight]{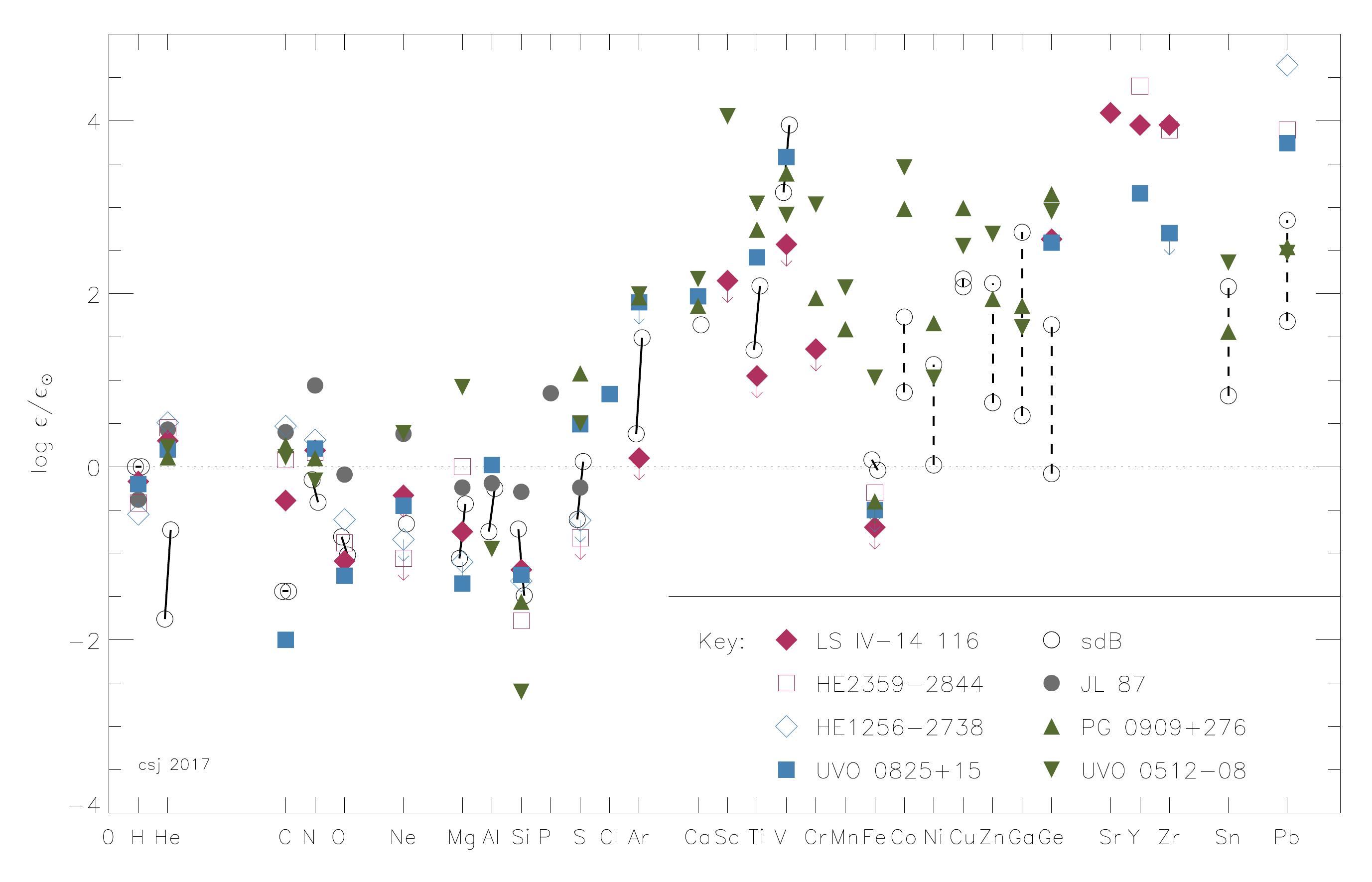}
\caption{\textbf{Top:} The distribution of  chemically-peculiar, helium-rich and normal hot subdwarfs with effective temperature and surface gravity. The solid line shows representative positions for the theoretical zero-age helium main-sequence (HeMS: $Z=0.02$) and zero-age horizontal branch (ZAHB). The observed data are from this work, \citet{naslim11,naslim13,jeffery17a} and \citet{nemeth12}. \textbf{Bottom:} Surface abundances of super metal-rich hot subdwarfs, including the pulsating stars LS IV\protect$-14^{\circ}116$ and UVO 0825+15, and the two stars considered in this paper; \uvstar\ and \pgstar. Abundances are shown relative to solar values (dotted line). Mean abundances and ranges for the helium-rich subdwarf JL 87 \protect\citep{ahmad07} and for normal subdwarfs are also shown. The latter are shown by connected open circles as (i) \protect$Z\leq26$ (solid lines): the average abundances for cool and warm sdBs \protect\citep{geier13} and (ii) \protect$Z\geq27$ (broken lines): the range of abundances measured for five normal sdBs from UV spectroscopy \protect\citep{otoole06}.}
\label{fig:teff_logg}
\label{fig:abs}
\end{figure*}

Figure \ref{fig:abs} shows the surface chemical abundance data for several metal-rich subdwarfs from previous analyses, as well as for the stars studied in this paper, as measured relative to the standard solar surface composition \citep{asplund09}. It shows the general upwards trend for heavier metals to be more enriched. The linked circles showing the abundance ranges for `normal' hot subdwarfs illustrate this.

In both \pgstar\ and \uvstar, all observed elements with $Z\ge16$ are enriched compared to solar. Elements with $Z<16$ are either roughly solar or, in the case of silicon, strongly subsolar. Whilst most hot subdwarfs show roughly solar surface nitrogen, the majority are carbon poor. In contrast, \pgstar\ and \uvstar\ are  carbon rich relative to other hot subdwarfs. 

	Although the majority of both normal sdB stars and peculiar subdwarfs have an iron abundance close to solar, \uvstar\ is detectably iron-rich, with $\rm [Fe] = 1.070\pm0.040$. We postulate that this is either due to iron stratification in the atmosphere or to the progenitor composition. 

    Figure \ref{fig:teff_logg} shows the position in \teff\ and \lgcs\ of both stars in this analysis, alongside other peculiar subdwarfs and the populations of known H-rich and He-rich subdwarfs.  We can see that both stars sit near the zero-age He main sequence, similar to other peculiar sdB stars. 
    
	We detected significant inconsistencies in several elements in each star. Tables \ref{table:pg_abun}\ and \ref{table:uvo_abun} summarise these discrepancies, and a regression fit was performed on these data. We found no statistically significant trends in the discrepancies of \pgstar\ or \uvstar.
	In the absence of any evidence for stratification in either star, we conclude that it is likely that both stars are intrinsically enriched with heavy elements to a degree rarely seen in hot subdwarfs. The question of why this subcategory of sdO/B stars is so rich in heavy metals remains open, and requires further thought. We have not ruled out the possibility of stratification by radiative levitation, but if this is to be observed, more sensitive methods are needed.

\section{Conclusion}
	We have carried out spectroscopic fine analyses of two helium-enriched hot subdwarfs, \pgstar\ and \uvstar. These are based on new ultraviolet spectra obtained with the {\it Hubble Space Telescope}, and a re-analysis of the optical spectra obtained by \cite{edelmann03th}. From the latter, we have remeasured \teff, \lgcs, and \nH and \nHe, as well as the abundances of C, N, Ne, S, Ar, Ca, Sc, Ti, V and Fe.  The new ultraviolet observations allowed us to measure abundances for  Cr, Mn, Fe, Co, Ni, Cu, Zn, Ga, Ge, Sn, and Pb in each star. In addition,   ultraviolet abundances of  lighter elements are typically in agreement with the optical measurements.
    
   We have demonstrated that both stars are enriched by 1.5 -- 3 dex in heavy metals ($Z\geq30$). The enrichment in zinc, gallium, tin and lead is consistent with 'normal' sdB stars \citet{otoole06}, and not as extreme as lead in \ledz, HE\,1256$-$2738, and HE\,2359$-$2844.
   Germanium, however, is more than 1 dex enriched compared with `normal' sdB stars, and similar to that seen in \ledz\ and \crimson. From these measurements, we confirm that both \pgstar\ and \uvstar\ have characteristics that make them distinct from both normal H-rich sdB and intermediate He-rich sdB stars.
   The additional iron-group elements measured from the ultraviolet confirm a picture of two extremely metal-enriched stars, even compared with `normal' sdB stars. \uvstar\ is possibly the most iron-rich hot subdwarf known to date, though \pgstar\ shares the low iron abundance characteristic of the lead-rich subdwarfs. 
   
   These extreme surface chemistries present a challenge for interpretation. We observed a number of discrepancies $>2\sigma$ between abundances measured from different wavelength regions, namely for S, Ca, and V in \pgstar, and C, S, Ca, Fe, Co, Ni, Cu, and Pb in \uvstar. Whether this is a consequence of fine-grained elemental stratification in the photosphere remains to be verified. 
   
   Meanwhile, the leading hypothesis for the extreme enrichment of all elements with $Z\geq18$ (except iron) remains that of a highly stratified surface in which selective radiative levitation has elevated abundant species into the line-forming region of the photosphere. These elements may contribute to missing opacity from the models, as suggested by the {\it IUE} spectrophotometry and {\it HST|/STIS} far-ultraviolet spectroscopy. Additional work is needed to identify and include the atomic data required. 
   
   
   There is also an urgent need to measure iron-group element abundances in other chemically peculiar hot subdwarfs, and especially in \crimson\ and \ledz\ where the high-resolution ultraviolet spectrum is easily accessible to {\it HST}/STIS.

\section*{Acknowledgments}

This paper is based on observations made with the NASA/ESA Hubble Space Telescope under program 13800, and recovered from the data archive at the Space Telescope Science Institute (STScI). STScI is operated by the Association of Universities for Research in Astronomy, Inc. under NASA contract NAS 5-26555.

The Armagh Observatory and Planetarium is funded by direct grant from the Northern Ireland Department for Communities.
CSJ acknowledges support from the UK Science and Technology Facilities Council (STFC) Grant No. 
ST/M000834/1. 

The authors acknowledge Aidan Kelly for work carried out for his senior sophister research project at Armagh Observatory and Trinity College Dublin in late 2015 and which provided the first indication of iron-group overabundances in \pgstar. 

The authors thank Ulrich Heber and Heinz Edelmann for making the optical spectra of \pgstar\ and \uvstar\ available. 

Some of the data presented in this paper were obtained from the Mikulski Archive for Space Telescopes
(MAST). STScI is operated by the Association of Universities for Research in Astronomy, Inc., under NASA 
contract NAS5-26555. Support for MAST for non-HST data is provided by the NASA Office of Space
Science via grant NNX09AF08G and by other grants and contracts.

This research has made use of the SIMBAD database, operated at CDS, Strasbourg, France.

This work has made use of the Vienna Atomic Line Database (VALD) database, operated at Uppsala University, 
the Institute of Astronomy of the Russian Academy of Sciences in Moscow, and the University of Vienna, 
the Atomic Line List, hosted by the Department of Physics and Astronomy, University of Kentucky,
and the National Institute of Standards and Technology (NIST) Atomic Spectra Database, which is hosted by the U.S. Dept of Commerce.


\bibliographystyle{mnras}
\bibliography{Leti}


\bsp	
\label{lastpage}
\end{document}